\documentclass[a4paper,11pt]{article}
\pdfoutput=1 

\usepackage{jinstpub} 

\usepackage{orcidlink}

\usepackage{lineno}
\usepackage{xcolor}

\title{\boldmath Radiation tolerance tests on key components of the ePIC-dRICH readout card}

\author[d,n,*]{S.~Geminiani\orcidlink{0009-0004-1673-9632},\note[*]{Corresponding author.
\\\\
This is the Accepted Manuscript version of an article accepted for publication in Journal of Instrumentation. Neither SISSA Medialab Srl nor IOP Publishing Ltd is responsible for any errors or omissions in this version of the manuscript or any version derived from it. The Version of Record is available online at DOI: 10.1088/1748-0221/21/04/C04039.}}

\author[d,n]{B.R.~Achari\orcidlink{0009-0001-1469-4117},}
\author[d,n]{N.~Agrawal\orcidlink{0000-0003-0348-9836},}
\author[h,q]{M.~Alexeev\orcidlink{0000-0002-7306-8255},}
\author[h,q]{C.~Alice\orcidlink{0000-0001-6297-9857},}
\author[j]{R.~Ammendola\orcidlink{0000-0003-4501-3289},}
\author[d]{P.~Antonioli\orcidlink{0000-0001-7516-3726},}
\author[d]{C.~Baldanza,}
\author[f]{L.~Barion\orcidlink{0000-0002-2453-9989},}
\author[i]{A.~Biagioni\orcidlink{0000-0001-5820-1209},}
\author[b,r]{A.~Calivà\orcidlink{0000-0002-2543-0336},}
\author[a,m]{M.~Capua\orcidlink{0000-0002-2443-6525},}
\author[i]{F.~Capuani\orcidlink{0000-0003-2026-6335},} 
\author[i,t]{A.~Ciardiello\orcidlink{0000-0003-1903-4406},} 
\author[i,k]{E.~Cisbani\orcidlink{0000-0002-6774-8473},}
\author[h,q]{M.~Chiosso\orcidlink{0000-0001-6994-8551},}
\author[f]{M.~Contalbrigo\orcidlink{0000-0002-8612-7998},}
\author[h]{F.~Cossio\orcidlink{0000-0003-0454-3144},}
\author[h]{M.~Da Rocha Rolo\orcidlink{0000-0001-8518-3755},}
\author[b,r]{A.~De Caro\orcidlink{0000-0002-7865-4202},}
\author[b,r]{D.~De Gruttola\orcidlink{0000-0002-7055-6181},}
\author[h]{G.~Dellacasa\orcidlink{0000-0001-9873-4683},}
\author[d]{D.~Falchieri\orcidlink{0000-0002-0255-8097},}
\author[a,m]{S.~Fazio\orcidlink{0000-0002-4321-1946},}
\author[i]{O.~Frezza\orcidlink{0000-0001-8277-1877},}
\author[b,r]{N.~Funicello\orcidlink{0000-0001-7814-319X},}
\author[d,l]{M.~Garbini\orcidlink{0000-0002-3016-6373},}
\author[h,s]{N.~Jacazio\orcidlink{0000-0002-3066-855X},}
\author[i]{F.~Lo Cicero\orcidlink{0000-0002-9904-2619},}
\author[i]{A.~Lonardo\orcidlink{0000-0002-5909-6508},}
\author[f]{R.~Malaguti\orcidlink{0000-0001-9576-6428},}
\author[c]{F.~Mammoliti, \orcidlink{0000-0003-1951-4901},}
\author[i]{M.~Martinelli\orcidlink{0000-0002-2585-3005},}
\author[h]{M.~Mignone\orcidlink{0009-0006-2973-3103},}
\author[h]{C.~Mingioni,} 
\author[h]{M.~Nenni,} 
\author[c]{F.~Noto\orcidlink{0000-0003-2926-7342},}
\author[a,m]{L.~Occhiuto\orcidlink{0009-0008-5620-9862},}
\author[d]{A.~Paladino\orcidlink{0000-0002-3370-259X},}
\author[h,s]{D.~Panzieri\orcidlink{0009-0007-4938-6097},}
\author[i]{P.~Perticaroli\orcidlink{0009-0005-7516-9235},}
\author[f]{S.~Plavully\orcidlink{0009-0009-0474-3278},} 
\author[f,o]{L.~Polizzi\orcidlink{0009-0005-4121-6435},}
\author[i]{L.~Pontisso\orcidlink{0000-0001-7137-5254},}
\author[d]{R.~Preghenella\orcidlink{0000-0002-1539-9275},}
\author[d]{R.~Ricci\orcidlink{0000-0002-5208-6657},}
\author[d]{L.~Rignanese\orcidlink{0000-0003-3167-0309},}
\author[b,r]{C.~Ripoli\orcidlink{0000-0002-6309-6199},}
\author[i,t]{C.~Rossi\orcidlink{0000-0001-5716-1401},} 
\author[d,n]{E.~Rovati\orcidlink{0009-0002-5128-4587},}
\author[d]{N.~Rubini\orcidlink{0000-0001-9874-7249},}
\author[h,s]{M.~Ruspa\orcidlink{0000-0002-7655-3475},}
\author[f]{A.~Saputi\orcidlink{0000-0001-6067-7863},} 
\author[i]{F.~Simula\orcidlink{0000-0002-7955-1491},}
\author[f]{F.~Spizzo\orcidlink{0000-0002-9134-4487},} 
\author[h]{U.~Tamponi\orcidlink{0000-0001-6651-0706},} 
\author[a,m]{E.~Tassi\orcidlink{0000-0002-3335-6500},}
\author[d]{G.~Torromeo\orcidlink{0009-0006-3411-7284},}
\author[e,p]{C.~Tuvè\orcidlink{0000-0003-0739-3153},} 
\author[i]{G.M.~Urciuoli\orcidlink{0000-0002-7559-0127},} 
\author[g]{S.~Vallarino\orcidlink{0000-0001-6679-0728},}
\author[i]{P.~Vicini\orcidlink{0000-0002-4379-4563},}
\author[h]{R.~Wheadon\orcidlink{0000-0001-8533-2132}}

\affiliation[a]{INFN Gruppo Collegato di Cosenza, Ponte Pietro Bucci 31 C, Rende, Italy}
\affiliation[b]{INFN Gruppo Collegato di Salerno, Via Giovanni Paolo II 132, Fisciano, Italy}
\affiliation[c]{INFN Laboratori Nazionali del Sud, Via S. Sofia 62, Catania, Italy}
\affiliation[d]{INFN Sezione di Bologna, Viale C. Berti Pichat 6/2, Bologna, Italy}
\affiliation[e]{INFN Sezione di Catania, Via S. Sofia 64, Catania, Italy}
\affiliation[f]{INFN Sezione di Ferrara, Via Saragat 1, Ferrara, Italy}
\affiliation[g]{INFN Sezione di Genova, Via Dodecaneso 33, Genova, Italy}
\affiliation[h]{INFN Sezione di Torino, Via Pietro Giuria 1, Torino, Italy}
\affiliation[i]{INFN Sezione di Roma 1, Piazzale Aldo Moro 2, Roma, Italy}
\affiliation[j]{INFN Sezione di Roma 2, Via della Ricerca Scientifica 1, Roma, Italy}
\affiliation[k]{Istituto Superiore di Sanità, Viale Regina Elena 299, Roma, Italy}
\affiliation[l]{Museo Storico della Fisica e Centro Studi e Ricerche Enrico Fermi, Via Panisperna 89 A, Roma, Italy}
\affiliation[m]{Università della Calabria, Ponte Pietro Bucci 31 C, Rende, Italy}
\affiliation[n]{Università degli Studi di Bologna, Viale C. Berti Pichat 6/2, Bologna, Italy}
\affiliation[o]{Università degli Studi di Ferrara, Via Saragat 1, Ferrara, Italy}
\affiliation[p]{Università degli Studi di Catania, Via S. Sofia 64, Catania, Italy}
\affiliation[q]{Università degli Studi di Torino, Via Pietro Giuria 1, Torino, Italy}
\affiliation[r]{Università degli Studi di Salerno, Via Giovanni Paolo II 132, Fisciano, Italy}
\affiliation[s]{Università del Piemonte Orientale, Via del Duomo 6, Vercelli, Italy}
\affiliation[t]{Sapienza Università di Roma, Piazzale Aldo Moro 5, Roma, Italy}

\emailAdd{sandro.geminiani2@unibo.it}

\abstract{The dual-radiator RICH \textcolor{black}{(dRICH)} detector of the ePIC experiment will employ over 300000 SiPM pixels as photosensors, organized into more than 1000 Photon Detection Units. Each PDU is a compact module, approximately $5 \times 5 \times 12~\mathrm{cm^{3}}$ in size, including four custom ASICs connected to 256 SiPMs and an FPGA-based readout card (RDO) responsible for data acquisition and control. 

Considering the moderately harsh radiation environment expected in the dRICH detector, this study reports on proton irradiation tests performed on key components of the RDO card to assess their tolerance to cumulative Total Ionizing Dose (TID) and Single Event Effects (SEE). All tested components demonstrated radiation tolerance beyond the TID levels expected for the dRICH environment, with the exception of the \textcolor{black}{ATtiny817} microcontroller, which showed destructive failure. Furthermore, as expected, the observed Single Event Upset (SEU) rates call for appropriate mitigation strategies in the final system design.
}


\proceeding{Topical Workshop on Electronics for Particle Physics\\
Rethymno, Crete, Greece\\
6–10 October 2025}

\begin{document}
\maketitle
\flushbottom

\section{Introduction}
\label{sec:intro}
The ePIC (Electron–Proton–Ion Collider) experiment will be the first collision experiment to explore the physics program of the future Electron–Ion Collider (EIC)~\cite{Yellow_report}.
As part of the ePIC particle identification system, the dRICH detector will provide $\pi$/K/p separation from 3 GeV/$c$ up to 50 GeV/$c$ and electron identification up to 15 GeV/$c$ \cite{ULLRICH2022167041}. \textcolor{black}{The selected photosensors are $3 \times 3~\mathrm{mm^{2}}$ SiPMs and are organized into 1248 PDUs, each integrating the readout electronics and 256 SiPMs.}

\begin{wrapfigure}{r}{0.4\textwidth}
    \centering 
    \includegraphics[width=0.35\textwidth,origin=c,angle=0]{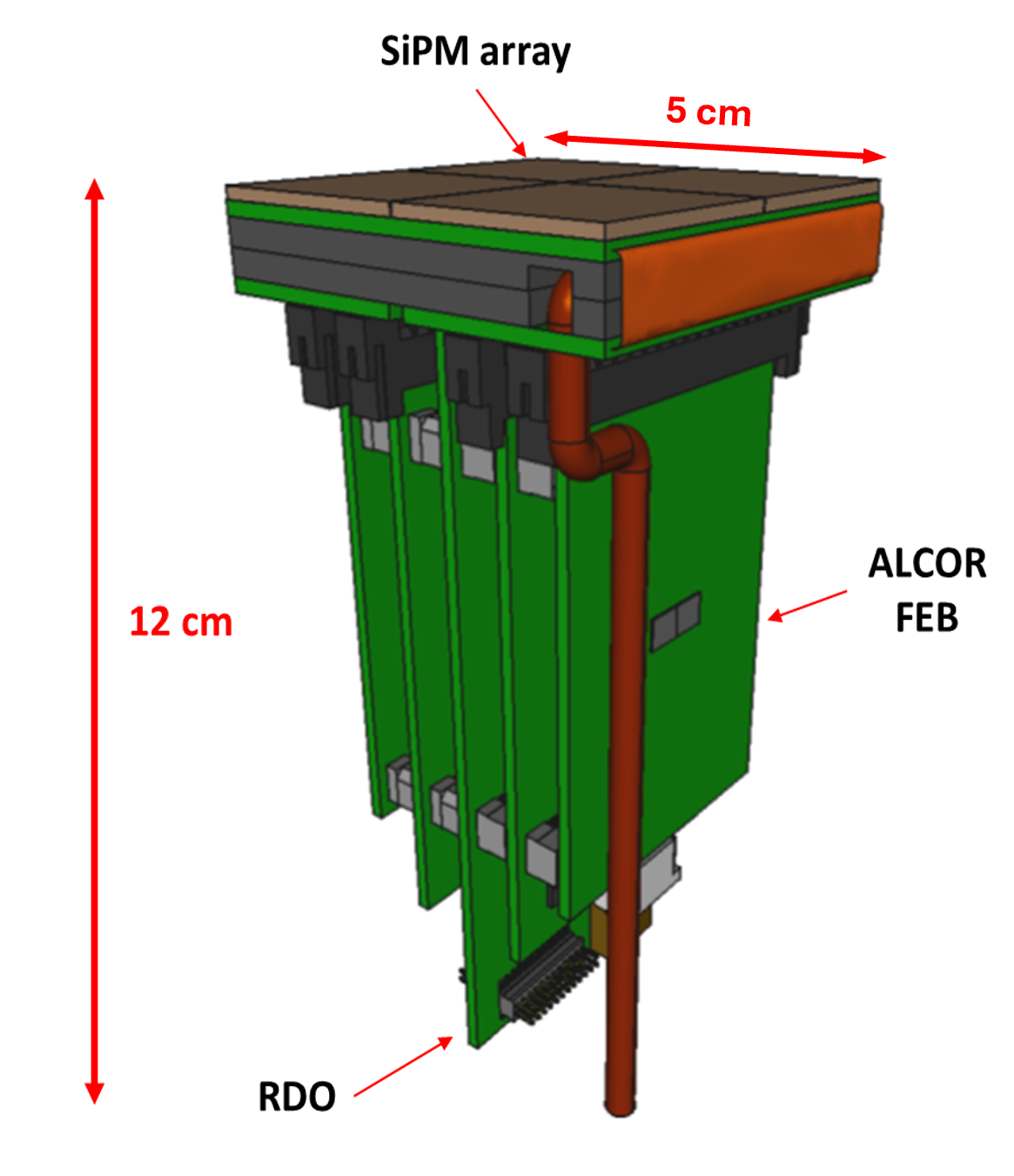}
    \caption{\label{fig:PDU} PDU design: SiPM arrays, the ALCOR ASIC Front-End Boards and the RDO readout card.}
\end{wrapfigure}
 
Within the PDU electronics region, the maximum hadron flux with energy greater than 20 MeV (relevant for SEU effects) is estimated to be $\phi_{5}(h > 20~\mathrm{MeV}) \simeq 700~\mathrm{Hz/cm^{2}}$. The expected total ionizing dose is $\mathrm{TID}_{5} \simeq 2.3~\mathrm{krad}$. Both values assume an integrated luminosity of 1000~fb$^{-1}$ and a safety factor of five. Cumulative TID effects and potential Single Event Latch-ups (SEL) may lead to permanent damage, while SEUs can corrupt device configuration or induce logic inversion in digital signals. 

The PDU, shown in figure~\ref{fig:PDU}, is the result of a collaborative effort among several INFN divisions, including the development of the SiPM sensor system, the ALCOR ASIC \cite{alcor} and the RDO readout card. In situ-annealing techniques have been intensively tested to partially recover SiPM performance after radiation-induced damage \cite{Achari_2025}. \textcolor{black}{Furthermore, irradiation tests of the RDO-selected components are initiated to assess their radiation tolerance and finalize the RDO layout. According to the test outcomes, device replacement will be considered and related SEU mitigation strategies will be developed.}

This contribution presents the results of the first two irradiation campaigns carried out at the Proton Therapy Center in Trento on key components of the RDO card.

\section{The RDO card}

\begin{figure}[htbp]
    \centering 
    \includegraphics[width=0.80\textwidth,origin=c,angle=0]{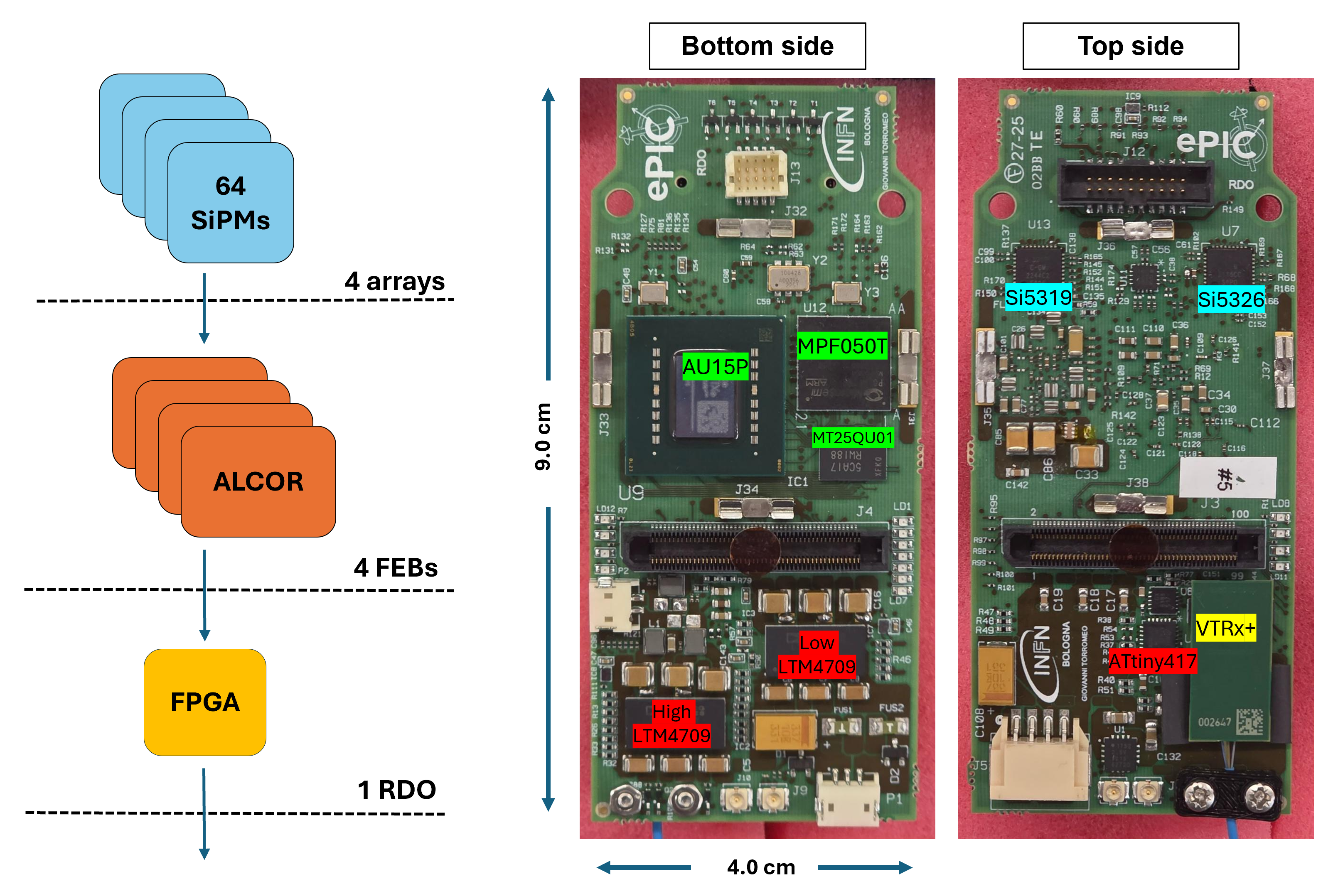}
    \caption{\label{fig:RDO} \textit{Left}: schematic of the PDU readout concept, showing the signal path from the 64-channel SiPM arrays through the ALCOR Front-End Boards to the AU15P SRAM-based FPGA hosted on the RDO card. \textit{Right}: layout of the RDO card, with the main components highlighted in different colors for each subsystem.}
\end{figure}

The PDU readout concept is illustrated on the left side of figure~\ref{fig:RDO}. It employs four 64-channel SiPM arrays as photosensors. Each array is coupled to a Front-End Board (FEB) hosting the ALCOR ASIC, which provides precise time stamping for each SiPM channel with a least significant bit (LSB) resolution of \textcolor{black}{40~ps}. All FEBs are connected to the RDO card via high-speed board-to-board connectors. The on-board FPGA on the RDO acts as a data concentrator for the ALCOR chips. The data-push architecture of ALCOR, together with the FPGA firmware design, enables a streaming readout scheme, transmitting data to the back-end system through a 10~Gb/s optical link.

The RDO must fit within the PDU module; therefore, its design is compact, while ensuring high-speed data communication and radiation tolerance. Figure~\ref{fig:RDO} (right) shows the RDO card, with its main components highlighted in different colors according to the corresponding subsystem:
\begin{itemize}
    \item \emph{FPGA subsystem}: an AU15P SRAM-based FPGA from AMD implements the high-speed readout of the differential data lines from the ALCOR chips. An MPF050T FLASH-based FPGA from Microchip reads and writes the AU15P Configuration RAM (CRAM) using an MT25QU01 FLASH memory from Micron Technology. The MPF050T operates as a scrubber for the AU15P CRAM to mitigate SEU occurrences.
    \item \emph{Clock network}: \textcolor{black}{two precision clock multipliers with jitter attenuation (Si5326 and Si5319, Skyworks Solutions) are used.} The Si5326 provides a 394~MHz clock to the ALCOR chips and for ALCOR–AU15P communication. The Si5319 generates the reference clock of the AU15P transceiver, using a 40~MHz on-board crystal oscillator \textcolor{black}{(Si511, Skyworks Solutions)} that also serves as the clock source for the MPF050T FPGA.  
    \item \emph{Power network}: \textcolor{black}{composed of two LTM4709 linear voltage regulators from Linear Technology; each regulator receives one of the two board power rails 1.4~V (Low-LTM4709) and 2.7~V (High-LTM4709).} Each LTM4709 integrates a triple-LDO circuit that generates three output voltages used for the board low-voltage networks. An ATtiny417 microcontroller ($\mu \text{C}$) from Microchip drives the power-up and power-down sequences of the board and monitors SEL conditions on each output rail.  
    \item \emph{Optical interface}: the VTRx+ module, developed at CERN, is a radiation-hard optical transceiver directly connected to the AU15P FPGA~\cite{vtrx+}.  
\end{itemize}

\section{Irradiation tests of key RDO components}
The first RDO prototypes were received in July~2025. Two irradiation campaigns were conducted in December~2024 and September~2025 to test several key RDO components hosted on commercial evaluation boards. The tests were performed at the Proton Therapy Center in Trento, using the experimental room instrumented by INFN-TIFPA. The proton beam characterization is described in~\cite{TOMMASINO201715}.

\subsection{Test setup instrumentation and methods}

Based on the available proton flux and beam spot size at different proton energies~\cite{TOMMASINO201715}, the optimal beam energy was selected to maximize the fluence on the device die while maintaining the SEU rate within controllable limits. Table~\ref{tab:flux} summarizes the proton beam configurations used for each Device Under Test (DUT). An ATtiny817 $\mu \text{C}$ was employed instead of the ATtiny417, as it provides the same package with larger memory resources.
\begin{table}[htpb]
    
    \caption{Irradiation parameters for each DUT (the commercial evaluation boards used are also listed)}
    \label{tab:flux} 
    \vspace{10pt}
    \centering
    \renewcommand{\arraystretch}{0.5}    
        \begin{tabular}{|c|l|c|c|} 
            \hline \rule[-3mm]{0mm}{0.3cm} \textbf{DUT} & \textbf{Evaluation board} & \textbf{Beam energy (MeV)} & \textbf{Flux (Hz/cm$^2$)} \\ 
            \hline \rule[-3mm]{0mm}{0.75cm} Si5326 & Skyworks Si5326 EVB & 100 & 10$^8$\\ 
            \hline \rule[-3mm]{0mm}{0.75cm} AU15P & Alinx AXAU15 & 70 & 10$^6$-10$^7$\\ 
            \hline \rule[-3mm]{0mm}{0.75cm} ATtiny817 & Microchip ATtiny817-XMINI & 100& 10$^7$-10$^8$\\ 
            \hline \rule[-3mm]{0mm}{0.75cm} LTM4709 & Analog Devices EVAL-LTM4709-BZ & 100 & 10$^7$-10$^8$ \\ 
            \hline 
        \end{tabular}
        
\end{table}

The TIFPA setup includes a \textcolor{black}{gaseous ionization detector} between the beam aperture and the DUT to measure total fluence and estimate the total ionizing dose with $\approx$ 5\% precision. A host PC controlled the DUT power supply and monitored current for SELs. 
The AXAU15 FPGA firmware monitored upsets in the Si5326 clock generator and in its own memory locations, using for CRAM the AMD Soft Error Monitor (SEM) IP core. The ATtiny817~$\mu \text{C}$ monitored its internal RAM and FLASH memories via the serial interface, while a Teensy 4.1 monitored the LDO voltage regulators.

Since most devices integrate multiple memory types, the Mean Time Between Failures (MTBF) for each DUT was estimated using the following equation:
\begin{equation}
    \mathrm{MTBF}_{\mathrm{dRICH}} = \left( N_{\mathrm{RDO}} \cdot N_{\mathrm{dev}} \cdot N_{\mathrm{bit}} \cdot \phi_{5} \cdot \sigma_{\mathrm{SEU}} \right)^{-1},
    \label{eq:MTBF}
\end{equation}
where $N_{\mathrm{RDO}}$ is the total number of dRICH RDO cards, $N_{\mathrm{dev}}$ the number of identical devices of the same technology integrated on the card, $N_{\mathrm{bit}}$ the maximum number of exposed bits, and $\sigma_{\mathrm{SEU}}$ the estimated SEU cross-section. Therefore, the $\mathrm{MTBF}_{\mathrm{dRICH}}$ value represents the average time interval between two consecutive SEUs in the entire dRICH system.

\subsection{Results}

In December~2024, the AU15P FPGA, the Si5326 clock multiplier and the ATtiny817~$\mu \text{C}$ were irradiated. A $\mathrm{TID}=49~\mathrm{krad}$ was accumulated on the Si5326 device, with no evidence of SEL or permanent damage. Using an IPbus connection~\cite{IPbus} between the AXAU15 board and the host PC, a SEU monitor was implemented to verify the configuration memory and check signals associated with output clock stability. \textcolor{black}{A $\sigma_{\mathrm{SEU}} = (1.8 \pm 0.4)\cdot10^{-14}~\mathrm{cm^2/~bit}$ was estimated for its configuration memory and an $\mathrm{MTBF}_{\mathrm{dRICH}} = (4.4 \pm 1.0)~\mathrm{h}$ was derived, considering both Skyworks Solutions devices hosted on the board.}
No clock-flag alerts were observed for the generated 394~MHz clock. Additionally, a period-jitter measurement was performed with an oscilloscope operating at a 40~GHz sampling frequency.
The measured RMS jitter remained stable, with a maximum deviation of approximately 0.1~ps.
The ATtiny817 device ceased functioning at \textcolor{black}{$\mathrm{TID} = 28~\mathrm{krad}$}, without evidence of SELs. \textcolor{black}{Considering its SRAM memory, a $\sigma_{\mathrm{SEU}} = (3.2 \pm 0.7)\cdot10^{-14}~\mathrm{cm^2/~bit}$ and an $\mathrm{MTBF}_{\mathrm{dRICH}} = (4.8 \pm 1.1)~\mathrm{h}$ were estimated. Since no SEUs were detected for its FLASH memory, 95\%~C.L. limits were computed as explained in~\cite{ESCC25100}: $\sigma_{\mathrm{SEU}} < 2.4\cdot10^{-16}~\mathrm{cm^2/~bit}$ and $\mathrm{MTBF}_{\mathrm{dRICH}} > 42~\mathrm{h}$.}
For the AU15P FPGA, a total dose of $\mathrm{TID} = 6.2~\mathrm{krad}$ was accumulated without any observed SELs or damage. 
The FPGA was programmed for self-detection of SEUs and via the IPbus link it was possible to monitor flip-flops, CRAM, and Block RAM (BRAM). The software on
the host PC monitored memory locations for errors, applying the appropriate corrective actions.
\textcolor{black}{The values of $\sigma_{\mathrm{SEU}}$ and $\mathrm{MTBF}_{\mathrm{dRICH}}$ are reported in table \ref{tab:AU15P_cross}.}

\begin{table}[!h]
    \caption{Estimated values of $\sigma_{\mathrm{SEU}}$ and $\mathrm{MTBF}_{\mathrm{dRICH}}$ for each AU15P resource type.}
    \label{tab:AU15P_cross} 
    \vspace{10pt}
    \centering
    \renewcommand{\arraystretch}{0.5}    
        \begin{tabular}{|c|c|c|} 
            \hline \rule[-3mm]{0mm}{0.3cm} \textbf{Resource type} & \textbf{$\sigma_{\mathrm{SEU}}$ (cm$^2$/bit)} & \textbf{$\mathrm{MTBF}_{\mathrm{dRICH}}$ (min)} \\ 
            \hline \rule[-3mm]{0mm}{0.75cm} CRAM & $(2.3 \pm 0.3)\cdot10^{-16}$ & $(2.4 \pm 0.3)$\\
            \hline \rule[-3mm]{0mm}{0.75cm} BRAM & $(1.8 \pm 0.2)\cdot10^{-15}$ & $(2.1 \pm 0.3)$\\
            \hline \rule[-3mm]{0mm}{0.75cm} Flip-flop & $< 4.3\cdot10^{-14}$ (95\%~C.L.)  & $> 2.8$ (95\%~C.L.)\\ 
            \hline
        \end{tabular}

\end{table}

\begin{figure}[htbp]
    \centering
    \includegraphics[width=0.87\textwidth]{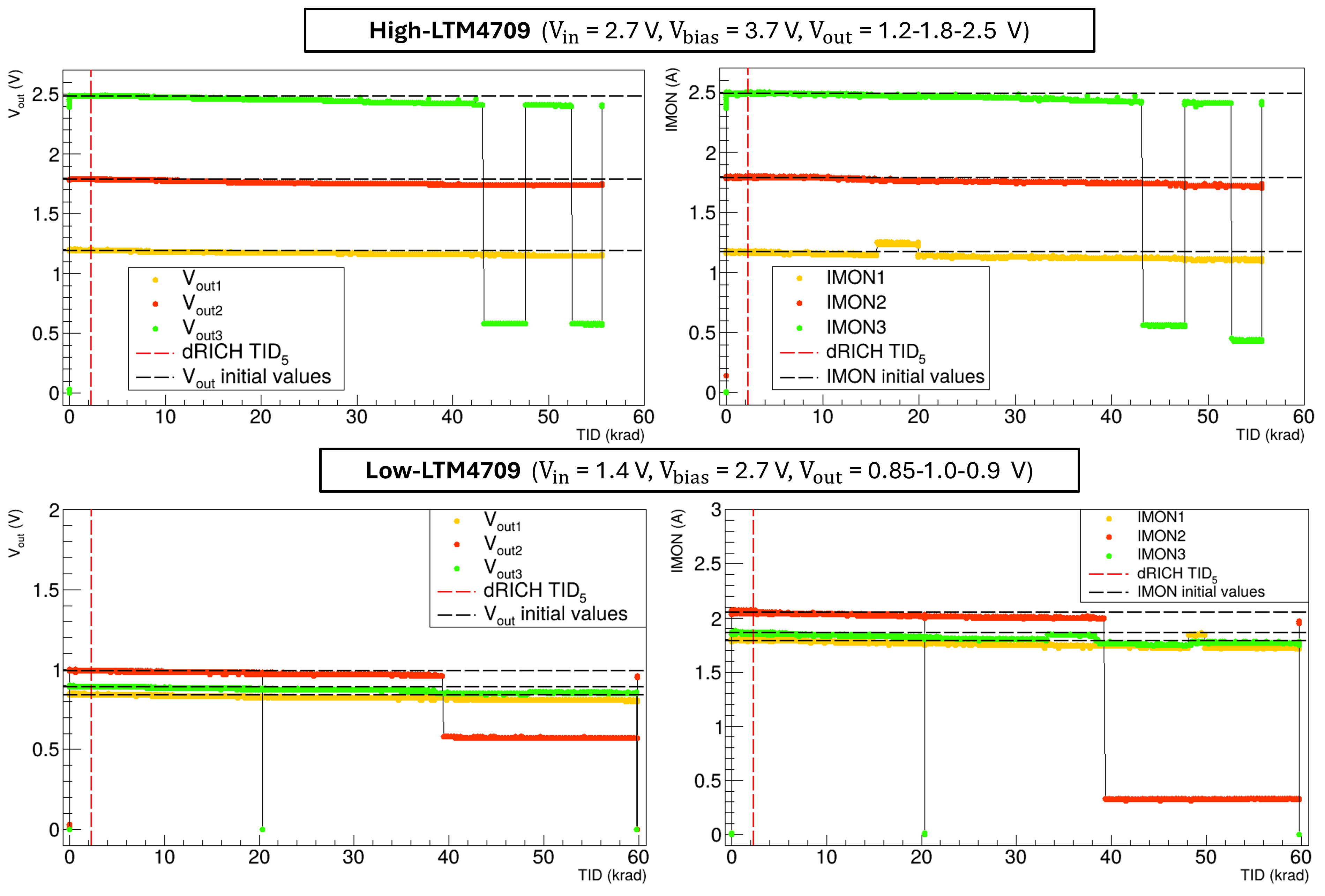}
    \caption{\label{fig:LTM}
    \textit{Top:} $V_{\mathrm{out}}$ and IMON as a function of TID for the High-LTM4709 configuration.
    \textit{Bottom:} $V_{\mathrm{out}}$ and IMON as a function of TID for the Low-LTM4709 configuration.}
\end{figure}

In September~2025, two LTM4709 devices were irradiated. The LTM4709 tests were carried out under two voltage configurations and power consumptions corresponding to the RDO operating configuration. 
The output voltages and current monitor (IMON) signals were monitored. The Power-Good (PG) signals, which provide alerts for $V_{\mathrm{out}}$ faults, were used as interrupt triggers to the $\mu \text{C}$ to detect SEU occurrences. In this context, a SEU was defined as a transient glitch in $V_{\mathrm{out}}$, IMON or PG signals, corresponding to a transition of the PG signals.
Figure~\ref{fig:LTM} shows the $V_{\mathrm{out}}$ and IMON evolution as a function of the integrated dose up to approximately 60~krad. A small decrease in both current and voltage was observed. At TID $\sim$ 60~krad a 5\% maximum decrease of the initial value was measured for both $V_{\mathrm{out}}$ rails and the IMON signals. No large variations were observed below $\mathrm{TID} = 20~\mathrm{krad}$. Occasional faults on the power rails, detected at higher doses, were recovered by toggling the Enable (EN) signals of each LDO circuit. Since no memory elements are exposed in this configuration, a value of $N_{\mathrm{bit}} = 1$ was used in equation~\ref{eq:MTBF}. \textcolor{black}{For doses up to 20~krad, a $\sigma_{\mathrm{SEU}} = (1.1 \pm 0.5)\cdot10^{-11}~\mathrm{cm^2/~bit}$ and an $\mathrm{MTBF}_{\mathrm{dRICH}} = (14 \pm 7)~\mathrm{h}$ were determined.} An additional test was also performed on five ATtiny817 $\mu \text{C}$s aligned along the beamline. The low radiation tolerance of the device was confirmed: as all five units ceased communication with the SEU monitor beyond $\mathrm{TID} > 8~\mathrm{krad}$, with a total accumulated dose of 31~krad across all devices. \textcolor{black}{For the SRAM memory, a higher $\sigma_{\mathrm{SEU}} = (5.6 \pm 1.1)\cdot10^{-14}~\mathrm{cm^2/~bit}$ and a lower $\mathrm{MTBF}_{\mathrm{dRICH}} = (2.8 \pm 0.5)~\mathrm{h}$ were estimated. Considering FLASH memory, new 95\%~C.L. limits were computed: $\sigma_{\mathrm{SEU}} < 4.2\cdot10^{-17}~\mathrm{cm^2/~bit}$ and $\mathrm{MTBF}_{\mathrm{dRICH}} > 240~\mathrm{h}$.}

\section{Conclusions and outlook}
\textcolor{black}{From the standpoint of radiation tolerance, the finalization of the RDO layout requires that all its components be validated through irradiation tests.} Following the first two test campaigns, all tested devices were qualified up to a TID exceeding $\text{TID}_5$, except the ATtiny817 $\mu$C, which failed destructively; alternative COTS solutions will replace the ATtiny417. SEU mitigation is required based on MTBF estimates: Skyworks jitter attenuators will be refreshed well below $\text{MTBF}_{\text{dRICH}}$, AU15P firmware will handle BRAM and flip-flop SEUs, and the MPF050T FPGA will implement CRAM scrubbing. The new $\mu$C will monitor V$_{\text{out}}$, IMON, and PG signals to control EN lines of both LTM4709 devices, in case of failure.

\textcolor{black}{The RDO card will be tested in future campaigns to validate the implemented SEU mitigation strategies, as well as the remaining selected components that have not yet been tested.}

\acknowledgments

This work is funded by the Commissione Scientifica Nazionale 3 (CSN3) of the Istituto Nazionale di Fisica Nucleare, Italy. This work is supported by the U.S. Department of Energy, Office of Science, Office of Nuclear Physics under the EIC project number JSA-22-R412967. The authors would like to thank the TIFPA staff for their help during the irradiation campaigns.

\bibliographystyle{JHEP}
\sloppy
\bibliography{biblio_1}









\end{document}